\documentclass[twocolumn]{aastex631}

\usepackage{graphicx} 
\usepackage{subfigure}
\usepackage{comment}
\usepackage{amsmath}
\usepackage[normalem]{ulem}
\usepackage{amsmath}

\usepackage{xcolor}

\begin{document}

\title{Follow-up of Neutron Star Mergers with CTA and  Prospects for Joint Detection with Gravitational-Wave Detectors} 

\author[0000-0002-9445-1405]{T. Mondal}
\affiliation{Department of Physics, Indian Institute of Technology Kharagpur, Kharagpur, West Bengal 721302 India}

\author[0009-0001-8084-0565]{S. Chakraborty}
\affiliation{Department of Physics, Indian Institute of Technology Kharagpur, Kharagpur, West Bengal 721302 India}

\author[0000-0001-9407-9845]{L. Resmi}
\affiliation{ Department of Earth $\&$ Space Sciences, Indian Institute of Space Science $\&$ Technology, Trivandrum 695547, India}

\author[0000-0003-1071-5854]{D. Bose}
\affiliation{Department of Physics, Central University of Kashmir, Ganderbal, Jammu $\&$ Kashmir 191131, India}

\begin{abstract}
The joint gravitational wave (GW) and electromagnetic observations of the binary neutron star (BNS) merger GW170817 marked a giant leap in multi-messenger astrophysics. The extensive observation campaign of the associated Gamma-Ray Burst (GRB) and its afterglow has strengthened the hypothesis associating GRBs with BNS mergers and provided insights on mass ejection, particularly the relativistic outflow launched in BNS mergers. In this paper, we investigate the joint detection probabilities of BNS mergers by GW detectors and the upcoming ground-based very-high-energy (VHE) $\gamma$-ray instrument, the Cherenkov Telescope Array (CTA). Using an empirical relation that constrains the distance-inclination angle plane, we simulated BNS mergers detectable in the O5 run of the LIGO/Virgo/Kagra (LVK) network with $300$~Mpc BNS horizon. Assuming Gaussian structured jets and ignoring large sky localization challenges of GW detectors, we estimated VHE afterglow detection probability by CTA. We have explored the afterglow parameter space to identify conditions favourable for detecting synchrotron self-Compton emission by CTA. Our study reveals that events viewed at angles $\lesssim3$ times the jet core angle are detectable by CTA when the initial bulk Lorentz factor at the jet axis ranges between 100 and 800. We find high kinetic energy ($E_k>10^{50}$ erg), ambient density ($n_0>10^{-1}$ $cm^{-3}$), and energy content in non-thermal electrons significantly enhance the likelihood of CTA detection within 300 Mpc. The joint detection rate varies significantly with afterglow parameter distributions, ranging from $0.003$ to $0.5$ per year. 
\end{abstract}


\section{Introduction}
Gamma-ray bursts (GRBs) are exceptionally intense and luminous bursts of gamma-rays releasing enormous amounts of energy, ranging from $10^{48}-10^{53}$~ergs \citep{klebesadel1973observations, Costa:1997obd, Kulkarni:1999aa}. For decades, binary neutron star (BNS) or neutron star-black hole (NS-BH) mergers were considered the potential progenitors of short-duration GRBs \citep{paczynski1986gamma, eichler1989nucleosynthesis, paczynski1991cosmological, narayan1992gamma, fryer1999formation}. 

In August 2017, the first observational evidence supporting this hypothesis emerged from the detection of the BNS merger GW170817 by LIGO\footnote{\url{https://www.ligo.caltech.edu/page/ligo-scientific-collaboration}} and Virgo\footnote{\url{https://www.virgo-gw.eu/about/scientific-collaboration/}} \citep{ abbott2017gw170817,abbott2020gravitational}. This event was spatially and temporally coincident with the $\gamma$-ray burst GRB170817A, detected by the Fermi-GBM \footnote{\url{https://fermi.gsfc.nasa.gov/science/instruments/gbm.html}} and INTEGRAL SPI-ACS \footnote{\url{https://www.cosmos.esa.int/web/integral/instruments-spi}}  \citep{goldstein2017ordinary, Savchenko_2017}.  

Further, it was followed by a non-thermal broad-band afterglow detected by XMM-Newton and Chandra in X-rays \citep{d2018evolution, haggard2017deep, margutti2017electromagnetic, troja2017x} and by the Very Large Array (VLA), the Australia Telescope Compact Array (ATCA), the Giant Metre-wave Radio Telescope (GMRT), the High Sensitivity Array (HSA), and the European VLBI Network (EVN) in radio frequencies \citep{hallinan2017radio, kim2017alma, mooley2018mildly, mooley2018strong,mooley2018superluminal,alexander2017electromagnetic, ghirlanda2019compact}. 

This landmark joint detection has opened up a significant era in multi-messenger astrophysics and offered profound insights into understanding the physics of compact object mergers and GRBs. The gravitational wave signal contains information about the time of the merger, the location, distance, orbital inclination, orbital phase, orbital eccentricity, and component masses and spins \citep{cutler1994gravitational, balasubramanian1996gravitational}. Meanwhile, the electromagnetic signal probes the nature of the relativistic jet and the origin of the prompt and broad-band afterglow emission \citep{zhang2001high, granot2002off, berger2007prompt}. However, there were no detections of the afterglow above X-ray frequencies. 
Fermi LAT reported a flux upper bound of $4.5\times 10^{-10}$ erg cm$^{-2}$ s$^{-1}$ that corresponds to luminosity upper bound of $9.7\times 10^{43}$ erg s$^{-1}$ \citep{ajello2018fermi}, which is 5 order of magnitude less luminous than other LAT detected short GRBs (sGRBs).

The MAGIC telescope\footnote{\url{https://magic.mpp.mpg.de/}} and the HESS \footnote{\url{https://www.mpi-hd.mpg.de/hfm/HESS/}} telescope followed up the event for several days but found no significant TeV emission \citep{stamerra2022follow, abdalla2017tev, abdalla2020probing}. One of the reasons is the intrinsic faintness of the GeV/TeV afterglow of GRB170817A. The other reasons are technical, such as the large sky localization region of the GW event and the slow response time of VHE detectors. Since 2019, MAGIC and HESS have detected VHE afterglows from a few GRBs originating from the gravitational collapse of massive stars (collapsars) beyond hundreds of GeVs.

MAGIC detected GRB190114C at a redshift($z$) of 0.4245 from $T_{0}$+57s to $T_{0}$+15ks \citep{acciari2019magic}, and GRB201216C ($z= 1.1$) from $T_{0}+56s$ to $T_{0}+73.8$ks \citep{abe2024magic}. HESS detected TeV emission from GRB 180720B ($z=0.653$) between 10 to 12 hrs \citep{abdalla2019very} and 190829A ($z=0.0785$) between 4 to 56 hrs after the trigger \citep{hess2021revealing}. Recently, LHAASO \footnote{\url{http://english.ihep.cas.cn/lhaaso/}} has reported the detection of VHE photons beyond 10 TeV from GRB 221009A \citep{lhaaso2023very}, another collapsar GRB, within $2000$~s from Trigger time. 

Despite multiple efforts, no detections of short GRBs have been made so far in the VHE regime except the $3 \sigma$ upperlimit of TeV emission from GRB160821B by MAGIC \citep{acciari2021magic} . From December 2014 to February 2017, HAWC \footnote{\url{https://www.hawc-observatory.org/}} telescope searched for VHE afterglow emission from multiple sGRBs but did not see any significant signal \citep{dichiara2017search, galvan2019search}. Like HAWC, there is also LHAASO\citep{lhaaso2023very}, a water Cherenkov detector, with its high sensitivity and wide sky-coverage, it expects to detect emission from sGRBs.

Soon, the upcoming Cherenkov Telescope Array (CTA) will be capable of detecting gamma rays across a wide energy spectrum, ranging from 20 GeV to 300 TeV \citep{acharya2017science, bose2022ground} and an excellent angular resolution compared to any current imaging $\gamma$-ray telescope. The unprecedented sensitivity and large effective area of CTA will significantly enhance the number of detectable VHE afterglows. It will be able to access a larger universe volume with improved sensitivity and offer a better observational strategy with its wide field of view (FOV) and quick response time. 

Meanwhile, the next-generation GW detectors are expected to increase the horizon and improve the sky localization. This improvement will enable timely alerts to be sent to VHE observatories, providing an estimate to localize the source by properly slewing the VHE detectors towards the source in time.

In this paper, we have explored the favourable parameter space of TeV detections of BNS counterparts and the possible rate of such events. We have modelled VHE off-axis emission ($>$ 100 GeV) from sGRBs during its afterglow phase predicted the detection prospects of sGRB-GW events with CTA during the O5 run of the LVK network in the coming years. 

Joint detection of BNS mergers by CTA and LIGO will improve parameter estimation of the afterglow and enhance the prospect of understanding particle acceleration mechanisms and magnetic field generation in jets associated with BNS mergers.

TeV afterglows of GRB can result from synchrotron \citep{Sari_1998, kumar2015physics}, external Compton \citep{zhang2021external}, synchrotron self-Compton (SSC) \citep{sari2001synchrotron}, or hadronic processes \citep{razzaque2010leptonic}. In this paper, we consider the SSC process incorporating the Klein-Nishina cross-section \citep{nakar2009klein}. We include internal absorption due to pair production \citep{joshi2021modelling}. The SSC flux is further corrected for absorption by the extragalactic background light (EBL) \citep{dominguez2011extragalactic}. Instead of simulating gravitational wave (GW) events, we apply a detection threshold based on the single detector sensitivity for a BNS horizon at 300 Mpc corresponding to O5, resulting in a condition involving the luminosity distance and inclination angle. For the afterglow jet, we assume a Gaussian structure, as seen in GRB170817A.

We investigate the influence of the afterglow parameter space on the detectability of afterglows. Through afterglow SSC light curves, we study how various afterglow parameters affect the TeV detections in case of CTA. We investigate primarily the effect of extrinsic parameters such as the observer's viewing angle to the jet axis $\theta_v$, luminosity distance $d_L$, and intrinsic parameters such as jet core angle $\theta_c$, kinetic energy $E_k$, ambient density $n_0$, electron energy fraction $\epsilon_e$, and magnetic field energy fraction $\epsilon_B$.

This paper is organized as follows: in section \ref{section2}, we have discussed off-axis afterglow emission of BNS merger events from Gaussian jet structure and its light curve behaviour. Further, we calculated the synchrotron and SSC fluxes, along with the Compton Y parameter and the Klein-Nishina (KN) effect on high-energy SSC spectra, including pair production and EBL correction. Section \ref{CTA-TEV} discusses the characteristics of TeV detectable afterglow flux in the CTA band. In section \ref{sect4}, we have explored the possibility of joint detection with gravitational wave (GW) detectors, including LIGO's upcoming detection criteria (O5 run), joint detection criteria by both CTA and LIGO, characteristics of joint detection events, and studies of events detected by CTA but not by LIGO, culminating with the joint detection rate of BNS mergers. Finally, in section \ref{Conclusion}, we have summarized our paper.

\section{VHE emission from misaligned Gaussian jets} \label{section2}

The jet associated with BNS mergers launches an external shock which interacts with the ambient medium. Non-thermal electrons and the magnetic field generated in the shock downstream result in synchrotron and synchrotron self-Compton (SSC) radiation, which can be observed from radio to TeV $\gamma$-rays \citep{totani1998very, sari2001synchrotron, fan2008high, ghirlanda2010onset}. Numerical simulations have shown that the jet develops a structure due to its interaction with the merger ejecta \citep{xie2018numerical, gottlieb2018cocoon, lazzati2018late, nakar2018implications}. Afterglow modelling of GRB170817A shows that the jet energy and bulk Lorentz factor follow a Gaussian structure \citep{zhang2002gamma, kumar2003evolution, rossi2004polarization, lamb2017electromagnetic}. The jet structure plays a crucial role in shaping the afterglow emission and influences the detectable electromagnetic signature of BNS mergers at different viewing angles, i.e., the angle between the line of sight of the observer and the jet axis.

\subsection{Off-axis afterglow}\label{section_2_afterglow}

In this paper, we consider a Gaussian structured jet similar to GRB170817A. The angular structure of the jet is characterised by $\theta_{c}$, which decides the sharpness. The other parameters are $\epsilon_c$ and $\Gamma_c$, the kinetic energy and the bulk Lorentz factor of the jet at its axis.

We assume the below angular profile for the energy per unit solid angle $\varepsilon (\theta)$ and the bulk Lorentz factor $\Gamma(\theta)$, respectively, used in \cite{resmi2018low}. 
\begin{equation} \label{eqn1}
    \varepsilon (\theta)\ =\epsilon_{c}\: e^{-\frac{\theta^{2}}{2\theta_{c}^{2}}}
\end{equation}
\begin{equation}\label{eqn2}
    \Gamma_{0}(\theta) \beta_{0} (\theta)= \Gamma_{c} \beta_{c} \exp\left ( -\frac{\theta^{2}}{2\theta_{c}^{2}} \right )
\end{equation}

Assuming the total energy of the jet to be $E_k$, jet kinetic energy per solid angle at its core $\epsilon_{c}$ can be written as $\epsilon_{c}=\frac{E_{k}}{\pi \theta_{c}^{2}(1-e^{-\theta_{j}^{2}/\theta_{c}^2})}$, where $\beta_{c}$ is the normalized velocity of the jet core corresponding to $\Gamma_c$.

The jet decelerates as it propagates and sweeps up the ambient material upstream. The kinetic energy is converted as the thermal energy of the shock downstream, leading to an evolution of the bulk Lorentz factor. The initial bulk Lorentz factor of the structured jet ranges from ultra-relativistic to sub-relativistic values. The adiabatic evolution of the jet bulk Lorentz factor in ultra-relativistic ($\Gamma \beta \gg 1$) phases is given by the Blandford-McKee self-similar solution \citep{blandford1976fluid}, $\Gamma \propto r^{-3/2}$ where $r$ is the radius of the blast wave from the centre of the explosion. In the non-relativistic regime ($\Gamma \beta \ll 1$), it is given as the Sedov-Taylor self-similar solution \citep{sedov1969similarity, taylor1950formation}, $\beta \propto r^{-3/2}$. Therefore, to simplify the calculations and to consistently account for the evolution in both regimes, we adopt an empirical form for the evolution following \cite{resmi2018low}, where $\Gamma(\theta,r) \beta(\theta,r) = \Gamma_{0}(\theta) \beta_{0}(\theta) \left ( \frac{r}{r_{dec}} \right )^{-3/2}$. Here, $r_{dec}$ is the deceleration radius at which the jet decelerates significantly. For the angular profile of $\varepsilon (\theta)$ and $\Gamma(\theta)$ we have defined in equations \ref{eqn1} and \ref{eqn2} receptively, $r_{\rm dec}$ does not depend on $\theta$ and is given by $(3\, \epsilon_{c}/\Gamma_{c}^{2}\beta_{c}^{2}\, n_{0}\, m_{p}\, c^{2})^{1/3}$, where $n_{0}$ is the density of the ambient medium, $m_{p}$ is the proton mass, and $c$ is the velocity of light.

To calculate the flux, we divided the jet structure into $m$ polar segments and $n$ azimuthal segments \citep{lamb2017electromagnetic}. The polar angle $\theta_{i}$, azimuthal angle $\phi_{k}$ that each segment makes with respect to the jet axis ranges from $0<\theta_{i}< 90^{\circ}$  and $0<\phi_{k}<2\pi$ respectively. However, in order to avoid numerical errors while integrating over the jet face, we restrict the $\theta_{i} \le 25^{\circ}$. This cut-off does not affect the total flux as emission from the low latitudes are not significant throughout when the light curve is above the detection threshold.

The fireball radius from where the emission originates corresponding to a given observed time depends on $\theta_i$ and $\phi_k$. To calculate this, we require the inclination angle of each segment with the observer's line of sight. For a segment at ($\theta_{i}$, $\phi_{k}$), the inclination angle is written as,
\begin{equation}
   \alpha_{incl}^{i,k}=\cos (\theta_{i}) \cos (\theta_{v}) + \sin (\theta_{i}) \sin (\theta_{v}) \cos (\phi_{k}).
\end{equation}

The relation between observer time $t_{\rm obs}$, shock radius from the central engine of the GRB $r$, and inclination angle $\alpha_{\rm inc}^{i,k}$ is given by,
\begin{equation}
    t_{obs}(r, \alpha_{inc}^{i,k}) = \frac{r}{\beta(r)c}\left [ 1-\beta(r) \cos \alpha_{inc}^{i,k} \right ].
\end{equation}

For a given $t_{\rm obs}$, we calculate the corresponding $r$ for each $\alpha^{i,k}_{inc}$ through interpolating the above equation. We integrate the flux from each segment arriving at the observer at a given time $t_{obs}$ to obtain the total flux. The off-axis flux from the segment at $\alpha^{i,k}_{inc}$ will be reduced by the doppler factor $a_{dop}(r,\alpha_{inc}^{i,k}) =\frac{1 - \beta(r)}{1 - \beta(r) \cos \alpha_{i,k}}$ such that, 
\begin{equation} \label{eqn_5}
\begin{split} 
     F^{\text{off}}_{\nu_{{i,k}}}  (t_{\rm obs}, \alpha_{\text{inc}}^{i,k}) &= a_{\text{dop}}^{3}(r,\alpha_{\text{inc}}^{i,k}) \times \cos \alpha_{\text{inc}}^{i,k} \\
     &\quad \times F^{\text{on}}_{\nu/a_{{i,k}}}(a_{\text{dop}}t_{\text{obs}}, \alpha_{\text{inc}}=0). 
\end{split}
\end{equation}
The frequency $\nu$ observed by the off-axis observer will correspond to the $\nu/a_{i,k}$ for the on-axis observer.

The on-axis flux received by an observer for each of the jet segments is given by,
\begin{equation}
   F^{\text{on}}_{\nu_{{i,k}}}  (t_{\rm obs}, \alpha_{\text{inc}}^{i,k}=0)= \frac{L_{\nu_{i,k}}}{4 \pi d_{L}^{2}} \times O_{fact}.
\end{equation}
Here, $L_{\nu_{i,k}}$ is the power emitted by the source at a specific frequency $\nu$ over an area of $4 \pi d_{L}^{2}$ across all the directions. $O_{fact}$ is the ratio 
$\frac{\Omega_{i,k}}{\Omega_{e,i,k}}$, where  $\Omega_{i,k}=\int_{\phi_{k-1}}^{\phi_{k}} d\phi\: \int_{\theta_{i-1}}^{\theta_{i}} d\theta \sin \theta$ is the solid angle subtended by a beamed element at a point on the central axis. $\Omega_{e, i,k}= {\rm max}\left [\Omega_{i,k}, 2 \pi(1-cos\frac{1}{\Gamma_{i,k}})  \right ]$.

\subsection{Calculation of Synchrotron flux}
To calculate the afterglow synchrotron light curve, we follow the standard assumptions as prescribed in \citet{wijers1999physical, piran1998spectra}. It is assumed that the non-thermal electron population is continuously injected inside the shock downstream following a power-law distribution in energy (equivalent to random motion Lorentz factor $\gamma_{e}$).  The minimum Lorentz factor $\gamma_{m}$ can be given by, $\gamma_{m} = \frac{m_{p}}{m_{e}}\left ( \frac{p-2}{p-1} \right ) \epsilon_{e}(\Gamma-1)+1$, where $m_{p}$ is the mass of the proton, $m_{e}$ is mass of the electron, $\epsilon_{e}$ is the fraction of thermal energy density of the downstream transferred to the electron. A uniform magnetic field is assumed behind the shock front, which can be represented as $B=(32 \pi m_{p} c)^{1/2} \epsilon_{B}^{1/2} n_{0}^{1/2} \Gamma$, where $\epsilon_{B}$ is the fractional thermal energy density transferred to the magnetic field. The synchrotron and SSC cooling becomes significant after $\gamma_c(t) = (1+z) \: \frac{6\pi m_{e}c}{\sigma_{T} (1+Y) B^{2} \Gamma t_{\rm obs}}$ where $Y$ is the Compton Y-parameter and $\sigma_{T}$ is Thomson scattering cross-section. We consider a maximum Lorentz factor $\gamma_{\rm max}$ for the electron distribution, representing the decrease in acceleration efficiency when the acceleration timescale exceeds the cooling timescale. $\gamma_{max}$ is given as $ \sqrt{6 \pi e\, e_{\rm acc} /\sigma_T B}$ by considering electron acceleration time scale at $e_{acc}=0.35$ \citep{zhang2020inverse}.

The resultant synchrotron light curve is characterised by time evolution of three break frequencies - minimum injection frequency $\nu_{m}(\gamma_{m})$, cooling frequency $\nu_{c}^{s}(\gamma_{c}^{s})$, and cut-off frequency $\nu_{max}(\gamma_{max})$ corresponding to the three electron Lorentz factors mentioned above. For an on-axis observer, the afterglow peak flux due to each of the jetted segments primarily depends on the total number of electrons $N_{e}$ present in that jet segment, which is proportional to $ \left ( \frac{n_{0}}{3} \right )r_{dec}^{3}\Omega_{i,k}$. The specific luminosity emitted by these electrons is distributed over a surface area of $d_{L}^{2}\Omega_{e, i,k}$. The spectral peak is decided by the power per unit frequency given by $P_{\nu,max} = m_{e}c^{2}\sigma_{T}B\sqrt{\Gamma^{2}+1}/3 q_{e}$. Hence, the on-axis synchrotron peak flux density for each jetted segment can be obtained as ---

\begin{equation}\label{eqn_ssc_peak}
 F_{\nu,max(i,k)}^{on}= P_{\nu,max} \times \left (\frac{n_{0}}{3}  \right )r^{3}O_{fact}\times \frac{1}{d_{L}^{2}}.   
\end{equation}

Accordingly, the on-axis flux from jetted segments can be calculated as ---

\begin{equation}\label{eqn_8}
 F_{syn(i,k)}^{on}= F_{max,syn(i,k)}^{on} \times f_{\nu,syn}. 
\end{equation}

Here $f_{\nu, syn}$ represents normalized synchrotron spectra which have two distinct spectral profiles based on the slow and fast cooling regimes and we have calculated this following \citet{Sari_1998} and \citet{zhang2018physics}. Using equation \ref{eqn_8}, the off-axis synchrotron flux can be calculated following equation \ref{eqn_5} in section \ref{section_2_afterglow}.

\subsection{Calculation of SSC flux}
Like synchrotron emission, the afterglow SSC light curve in the slow and fast cooling regime is also characterised by three characteristic break frequencies, $\nu_{mm}^{SSC}=2\nu_{m}\gamma_{m}^{2}$ , $\nu_{cc}^{SSC}=2\nu_{c}\gamma_{c}^{2}$ and $\nu_{mc}^{SSC}=\sqrt{\nu_{mm}^{SSC}\nu_{cc}^{SSC}}$. These break frequencies follow the convention $\nu_{a,b}^{SSC}= 2\nu_{a}\gamma_{b}^{2}$, where subscripts a,b= (m,c) denoting minimum injection and cooling frequency respectively. Hence, in the SSC regime, electron distribution will be in the slow cooling regime if $\nu_{mm}^{SSC}<\nu_{cc}^{SSC}$ and fast cooling regime if $\nu_{mm}^{SSC}>\nu_{cc}^{SSC}$. 

Likewise, the on-axis synchrotron flux (equation \ref{eqn_8}), for each of the jetted segments, the on-axis SSC flux can be calculated as,

\begin{equation} \label{eqn_9}
\begin{aligned}
     F_{SSC(i,k)}^{on} &= F_{max,syn(i,k)}^{on} \times (n_{0}r\sigma_{T}x_{0}) \times f_{\nu,SSC} \\
     &= F_{max,SSC(i,k)}^{on}\times f_{\nu,SSC}.
\end{aligned}
\end{equation}

Here the term $F_{max,SSC(i,k)}^{on}$ represents maximum peak flux due to the SSC effect, which accounts for synchrotron peak flux $F_{max, syn(i,k)}^{on}$ along with radius r of the blast from the centre of the explosion, number density $n_{0}$ of the ambient medium and Thomson scattering cross-section $\sigma_{T}$. We computed the on-axis SSC flux profile $f_{\nu, SSC}$ in the slow and fast cooling regime following \citet{sari2001synchrotron} and \citet{zhang2018physics}, where the entire SSC flux distribution can be calculated by integrating over synchrotron seed photon spectrum and the electron energy distribution. Additionally, we adopted a value of constant parameter $x_{0}$ as $0.5$ \citep{sari2001synchrotron}. 

The value of SSC peak flux $F_{max, SSC(i,k)}^{on}$ provides information about the relative importance of the synchrotron-self Compton process compared to synchrotron emission, which helps to determine how much energy is transferred from low-energy synchrotron photons to high-energy SSC photons. This affects the cooling rates of electrons and also the shape of the electron energy distribution over time. Thus, using the on-axis SSC flux from equation \ref{eqn_9}, we further compute the off-axis SSC flux following equation \ref{eqn_5}. As we are modelling afterglow emission of very high energy BNS merger events, we did not incorporate the effect of self-absorption frequency into our afterglow model.

\subsection{Compton Y parameter}
We have consistently extended our GRB afterglow model by incorporating the effect of the SSC process along with the synchrotron effect, which plays a crucial role in the electron cooling process. Initially, we explained the SSC effect by introducing the Y parameter in the Thomson regime, which accounts for the relative importance of SSC to synchrotron power. The effect takes place when the energy of the incoming photons becomes less than the rest mass energy of the electron ($h\nu << m_{e}c^{2}$). Following \citet{jacovich2021modelling} and \citet{mondal2023probing}, the Y parameter in the Thomson regime can be written as,

\begin{equation}
    Y_{TH}=min\left \{ Y_{slow},Y_{fast} \right \}
\end{equation}
Here $Y_{slow}$ depend on shock microphysical parameters $\epsilon _{e}$, $\epsilon _{B}$ and observer time $t_{obs}$, whereas $Y_{fast}$ only depends on $\epsilon _{e}$, $\epsilon _{B}$ and this holds for both $Y>>1$ and $Y<<1$. In this Thomson scattering process, interaction cross-section is insensitive to electron or photon energy, which is an elastic scattering process. In presence of Y parameter, electron cooling increases significantly and reduced cooling Lorentz factor such that the effective cooling Lorentz factor $\gamma_{c}$ becomes, $\gamma_{c}=\gamma_{c}^{s}/(1+Y_{TH})$. Hence, the overall synchrotron flux also decreases with modifying cooling frequency to $\nu_{c}$ as, $\nu_{c}$=$\nu_{c}^{s}/(1+Y_{TH})^{2}$. Thus, the SSC effect significantly modifies the synchrotron spectrum due to Thomson scattering.

\subsection{KN effect on high energy SSC spectra}

In the previous section, we have considered the SSC scattering under the assumption that Klein-Nishina (KN) effects are negligible. However, the energies of sub-TeV photons when comparable to or larger than electron rest mass energy $\gamma m_{e} c^{2}$ in the electron centre of mass frame, there is a significant transition of scattering cross-section from Thomson regime to KN regime \citep{nakar2009klein} and emissivity beyond this regime is significantly reduced compared to the Thomson regime. Above the KN limit, the energy gain of each photon in each scattering becomes constant. Thus, this transition results in the suppression of high-energy upscattered photons and affects the energy distribution of cooling electrons by modifying the high-energy spectral breaks.

The ratio of SSC to synchrotron emissivity depends on $\gamma$, thus the $Y$ parameter becomes
\begin{equation}
    Y(\gamma)=\frac{P_{SSC}(\gamma)}{P_{syn}(\gamma)}.
\end{equation}
We calculate the Compton parameter $Y(\gamma)$ as a function of electron energy, incorporating potential corrections due to KN effects. In the presence of KN effects, a set of characteristic Lorentz factors appears, which further modifies the slow-cooling or fast-cooling spectrum. Apart from typical Lorentz factors $\gamma_c$ and $\gamma_m$, following \citet{nakar2009klein} we define a critical Lorentz factor $\hat{\gamma}_{KN}$, given by 
\begin{equation}
    \hat{\gamma}_{KN}(\gamma) = \frac{\Gamma m_e c^2}{h \nu_{\text{syn}}(\gamma) (1 + z)}
\end{equation}
which represents the maximal Lorentz factor of an electron that can efficiently upscatter a synchrotron photon emitted by electrons with Lorentz factor $\gamma$, and beyond $\hat{\gamma}_{KN}$ there will be no upscattering as it is going above KN limit. Here $\nu_{syn}(\gamma)$ is the synchrotron frequency for an electron with Lorentz factor $\gamma$, which is given by $\nu_{\text{syn}}(\gamma) = \frac{\Gamma \gamma^2 q_e B}{2 \pi m_e c (1 + z)}$. Thus KN correction introduces additional critical Lorentz factors beyond \(\gamma_m\) and \(\gamma_c\), those are $\hat{\gamma}_m=\hat{\gamma}(\gamma_m)$ and $\hat{\gamma}_c=\hat{\gamma}(\gamma_c)$. Depending upon these newly induced critical Lorentz factors, new break frequencies can be introduced, $\hat{\nu}_m=\nu_{syn}(\hat{\gamma}_{m})$ and $\hat{\nu}_c=\nu_{syn}(\hat{\gamma}_{c})$. Moreover, there is another Lorentz factor $\gamma_0$, above which SSC cooling becomes ineffective, with $Y(\gamma_0) = 1$. Hence its corresponding break frequency becomes $\hat{\nu}_0=\nu_{syn}(\hat{\gamma}_{0})$. We further define another critical Lorentz factor $\gamma_{self}$, which can be defined as $\gamma_{self}=(B_{cr}/B)^{1/3}$. Here, $B_{cr}$ is the quantum critical field, and we used its numerical value $\sim$ 4.4 $\times 10^{13}$ Gauss in our calculation \citep{nakar2009klein}. The SSC spectrum varies in shape depending on the interaction between these different characteristic electron Lorentz factors and their corresponding frequencies. These additional frequencies further modify the predicted SSC flux levels along with its logarithmic correction terms. Although broadly, the spectral types can be categorized into slow (\(\gamma_m < \gamma_c\)) and fast (\(\gamma_c < \gamma_m\))\citep{Sari_1998} cooling regimes, the KN correction further modifies the spectral breaks, which we will discuss now.

In the slow-cooling regime, the synchrotron luminosity is primarily dominated by electron cooling Lorentz factor $\gamma_c$ and hence $Y(\gamma_c)$ represents the SSC luminosity to synchrotron luminosity ratio. Now, this regime can further be divided into two spectral categories \citep{nakar2009klein, yamasaki2022analytic}: the weak KN regime where \(Y(\gamma_c) \gtrsim 1\), and the strong KN regime where \(Y(\gamma_c) \lesssim 1\). Depending on the position of \(\gamma_c\), \(Y(\gamma_c)\) generally takes the following form \citep{nakar2009klein},

\begin{equation}
Y(\gamma_c) = Y(\hat{\gamma}_c) 
\begin{cases} 
1 & \gamma_c < \hat{\gamma}_c \\ 
\left( \frac{\gamma_c}{\hat{\gamma}_c} \right)^{\frac{p-3}{2}} & \hat{\gamma}_c < \gamma_c < \hat{\gamma}_m \\ 
\left( \frac{\gamma_c}{\gamma_m} \right)^{p-3} \left( \frac{\gamma_c}{\hat{\gamma}_m} \right)^{-\frac{4}{3}} & \hat{\gamma}_m < \gamma_c. 
\end{cases}
\label{Y_slow}
\end{equation}

The value of $Y(\hat{\gamma}_c)$ can be found by the normalization at $\gamma_c$:

\begin{equation}\label{Y_slwh}
Y(\gamma_c)[1 + Y(\gamma_c)] \approx \frac{\epsilon_e}{\epsilon_B} \left( \frac{\gamma_c}{\gamma_m} \right)^{2-p} \left( \frac{\min\{\gamma_c, \hat{\gamma}_c\}}{\gamma_c} \right)^{\frac{3-p}{2}}.
\end{equation}

We have obtained the value of $ Y(\gamma_c) $ numerically by substituting $\gamma_c = \frac{\gamma_{c}^{s}}{1+Y(\gamma_c)}$ and $\hat{\gamma}_c = \hat{\gamma_{c}^{s}} (1+Y(\gamma_c))^2$ into equations \ref{Y_slow} and \ref{Y_slwh}. Here, $\gamma_{c}^{s}$ and $\hat{\gamma}_{c}^{s}$ are defined as follows: $\gamma_{c}^{s} = \frac{6 \pi m_e c}{\sigma_T} \frac{(1 + z)}{\Gamma t B^2}$; $\hat{\gamma}_{c}^{s} = \frac{\Gamma m_e c^2}{h \nu_c^s}$ with $\nu_c^s = \frac{3}{4 \pi} \frac{q_e}{m_e c} \Gamma \frac{\gamma_c^2 B}{1 + z}$ where $h$ is the Planck constant. Invoking these substitution, equation \ref{Y_slow} and \ref{Y_slwh} modified as,

\begin{widetext}
\begin{equation}
Y(\gamma_c) = Y(\hat{\gamma}_c) 
\begin{cases} 
1 & \gamma_c < \hat{\gamma}_c \\ 
\left( \frac{\gamma_c^s}{\hat{\gamma}_c} \right)^{\frac{p-3}{2}} \left ( 1+Y(\gamma_c) \right )^{\frac{3-p}{2}} & \hat{\gamma}_c < \gamma_c < \hat{\gamma}_m \\ 
\left( \frac{\gamma_c^s}{\gamma_m} \right)^{p-3} \left( \frac{\gamma_c^s}{\hat{\gamma}_m} \right)^{-\frac{4}{3}} \left ( 1+Y(\gamma_c) \right )^{\frac{13}{3}-p} & \hat{\gamma}_m < \gamma_c. 
\end{cases}
\label{Y_slow_mod}
\end{equation}

and

\begin{equation}\label{Y_slwh_mod}
Y(\gamma_c)[1 + Y(\gamma_c)] \approx \frac{\epsilon_e}{\epsilon_B} \left( \frac{\gamma_c^s}{\gamma_m} \right)^{2-p}
\begin{cases} 
\left ( 1+Y(\gamma_c) \right )^{{p-2}} & \gamma_c < \hat{\gamma}_c \\ 
\left( \frac{\hat{\gamma}_c^s}{\gamma_c^s} \right)^{\frac{3-p}{2}}\left ( 1+Y(\gamma_c) \right )^{\frac{5-p}{2}} & \hat{\gamma}_c < \gamma_c. 
\end{cases}
\end{equation}
\end{widetext}

The slow cooling SSC spectrum follows the breaks as outlined by \citet{sari2001synchrotron} and as pointed out by \citet{nakar2009klein}, KN modification occurs at $\nu > \hat{\nu}_{cc}$ where $\hat{\nu}_{cc} = 2 \nu_c \hat{\gamma_c} \max{\{\gamma_c, \hat{\gamma_c}\}}$. Therefore, the KN-modified SSC spectrum becomes,

\begin{equation}
F_{\nu}^{\text{SSC}} \propto 
\begin{cases} 
\nu^{-p/2} & \nu_{cc}^{SSC} < \nu < \hat{\nu}_{cc} \\
\nu^{-(p-1)} & \hat{\nu}_{cc} < \nu < \hat{\nu}_{cm0} \\
\nu^{-(p+1)/2} & \hat{\nu}_{cm0} < \nu < \hat{\nu}_{cm} \\
\nu^{-(p+\frac{1}{3})} & \hat{\nu}_{cm} < \nu
\end{cases}
\end{equation}

where $\nu_{cc}^{SSC}=2\nu_{c}\gamma_{c}^{2}$, $\hat{\nu}_{cc}=2\nu_c \hat{\gamma}_c^2$, $\hat{\nu}_{cm0}= 2\nu_c \hat{\gamma}_c \min\{\hat{\gamma}_m , \gamma_0\}$, $\hat{\nu}_{cm}=2\nu_c \hat{\gamma}_c \gamma_m$. Now, only when $\gamma_c < \hat{\gamma}_c$ then the first segment of the SSC spectrum is observed, while the third segment appears if $\gamma_0 < \hat{\gamma}_m$. In cases where $\gamma_c < \hat{\gamma}_c$, the SSC peak occurs at $\nu_{\text{peak}}^{\text{SSC}} \approx 2\nu_c \gamma_c^2$. However, when $\gamma_c > \hat{\gamma}_c$, the SSC peak shifts to $\nu_{\text{peak}}^{\text{SSC}} \approx 2\nu_c \gamma_c \hat{\gamma}_c$ \citep{nakar2009klein}.

Fast cooling regime ($\gamma_c < \gamma_m$) is mainly categorized into three types: the weak KN regime, where $\gamma_m < \widehat{\gamma_m}$, the strong KN regime, where $\gamma_m > \widehat{\gamma_m}$ and the regime when $\gamma_m = \widehat{\gamma_m}$. Depending upon the strength of the KN effect and further appearance of secondary Lorentz factors $\gamma_{0}$ and $\gamma_{self}$, along with $\gamma_{m}$, the strong KN regime is further divided into three subclasses: regime I - $\gamma_{m}>\gamma_{self}>\gamma_{0}$, regime II- $\gamma_{m}>\gamma_{0}>\gamma_{self}$ and regime III- $\gamma_{0}>\gamma_{m}>\gamma_{self}$. For each of these cases, weak KN, strong KN and when $\gamma_{m} = \hat{\gamma_{m}}$, we first calculated $Y(\gamma_{m})$ and consecutively the SSC spectrum following the analytical solution provided by \citet{nakar2009klein}.

For accurate calculation of the SSC flux, we further account for optical depth due to pair production in the fireball and attenuation due to extragalactic
background light during the propagation (section \ref{Pair_EBL}), both of which significantly impact the intrinsic flux and temporal evolution of the afterglow. 

\subsection{Pair production and EBL correction} \label{Pair_EBL}

TeV photons from distant extragalactic sources interact with low-frequency photons (optical/UV/near and far IR) and produce electron-positron ($\gamma \gamma \rightarrow e^{-}e^{+}$) pair \citep{gould1967pair}. Consequently, these VHE gamma ray photons are absorbed. However, this interaction can occur in two distinct places. Firstly, at the source, these high-energy photons interact with low-energy photons and produce pairs. The optical depth of pair production measures the likelihood of those high-energy photons at the source. By calculating gamma ray flux at the source, we have taken care of the attenuation due to this pair production absorption. Hence, the intrinsic spectrum can be modified as $F_{int}' = F_{int}/ ( 1+ \tau_{\gamma \gamma })$ \citep{zhang2021external}, where $\tau_{\gamma \gamma }$ is the optical depth of photon opacity. Here, we have calculated energy-dependent optical depth $\tau_{\gamma \gamma }(E_{\gamma})$ following the methodology of \citet{gould1967pair} and \citet{joshi2021modelling}. Secondly, while passing through extragalactic space, gamma rays can again interact with EBL (Extragalactic background light) photons \citep{gould1966opacity, stecker1992tev, gilmore2012semi, ackermann2012imprint}. These low-frequency photons mainly come from the dust of the host galaxy \citep{finke2010modeling}. As a result, the spectrum we observe on Earth is modified by this EBL absorption. This EBL attenuation factor depends on both energy and redshift \citep{finke2010modeling}. Hence, the higher the redshift, the higher the attenuation of the observed flux.

\section{TeV Afterglow and CTA detectabiliy}\label{CTA-TEV}
In this section, we study TeV afterglows resulting from synchrotron photons up-scattered by the same relativistic electrons that emitted them through synchrotron-self-Compton (SSC) emission. In the previous section, we describe the theoretical model used for predicting TeV light curves from off-axis Gaussian jets. In this section, we investigate the parameter space of the model favouring CTA detection and have also explored the dependency of the temporal evolution of afterglow lightcurve on the observer viewing angle $\theta_{v}$ and jet core angle $\theta_{c}$.

\subsection{Light-Curve Characteristics} \label{LCs}

As the shock decelerates, relativistic beaming becomes less severe, alleviating doppler de-boosting. Consequently, the afterglow emission becomes visible to observers away from the jet core. For an observer with the jet axis aligned to their line of sight ($\theta_{v}=0$, on-axis), the light curve typically peaks early, around the onset of the blastwave deceleration. For an off-axis observer, the light curve peak is delayed to $\Gamma (\theta_v) \sim 1/\theta_{v}$. The peak flux is diminished significantly in this case due to Doppler de-boosting (see equation \ref{eqn_5}). Consequently, the off-axis observer as $\theta_{v}$ increases sees a lower flux and a delayed peak in the light curve.

\begin{figure}[ht] 
    \centering
    \includegraphics[width=\linewidth]{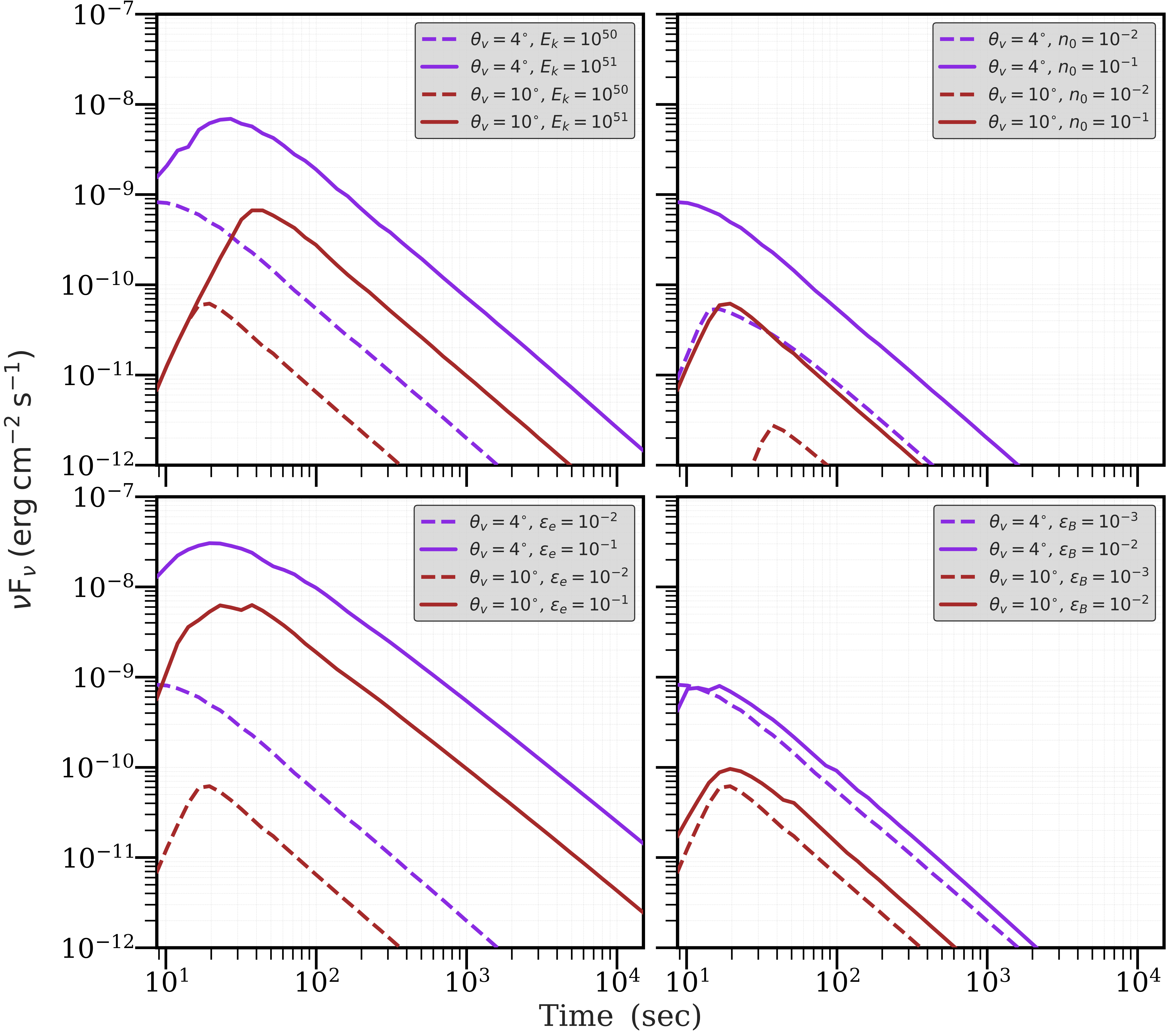}
    \caption{The SSC flux as a function of time is plotted for different combinations of afterglow extrinsic and intrinsic parameters. In each panel, jet structure parameter $\theta_{c}$ is kept fixed at $8^{\circ}$ and observer viewing angle $\theta_v$ is varied to $4^{\circ}$ and $10^{\circ}$. Along with $\theta_{v}$, another microphysical parameter, $E_{k}$ (upper-left panel) or $n_{0}$ (upper-right panel) or $\epsilon_{e}$ (lower-left panel) or $\epsilon_{B}$ (lower-right panel), is varied keeping all other parameters constant. The variation range of $E_{k}$, $n_{0}$, $\epsilon_{e}$, $\epsilon_{B}$ are --- $10^{50}$ and $10^{51}$ erg; $10^{-2}$ and $10^{-1}$ $cm^{-3}$, $10^{-2}$ and $10^{-1}$; and $10^{-3}$ and $10^{-2}$ respectively. Whereas the fixed value of $E_{k}$, $n_{0}$, $\epsilon_{e}$, $\epsilon_{B}$, $\Gamma_c$ are $10^{50}$ erg; $10^{-1}$ $cm^{-3}$, $10^{-2}$, $10^{-3}$ and $650$ respectively. Considering observer frequency of $250$ GeV and luminosity distance $d_{L}=250$ Mpc, pair production optical depth and the EBL attenuation factor is calculated. \label{CTA_LC_N}}
\end{figure}

To obtain the detectability of TeV afterglows by CTA, it is crucial to estimate the time scale beyond which afterglow flux remains above the detector threshold. CTA can detect fainter sources with a longer integration time. We choose 250 GeV as a representative band and have simulated light curves in this frequency. We have taken the differential flux sensitivity of the CTAO-Northern\footnote{\url{https://www.ctao.org/for-scientists/performance}} array at energy 250 GeV to decide the detection threshold for the simulated light curves. We define a detection if at least one point in the simulated light curve exceeds the threshold. Specifically, for a lightcurve at 250 GeV, where the flux $f_{250} (t)$ varies as a function of time, a detection occurs if $f_{250} (t) > S(t)$, where $S(t)$ is the threshold corresponding to an integration time $t$. Given that the light curve decays as a power-law, it is reasonable to consider that at a given time $t$, integration for a duration $t$ is possible.

The peak flux and peak time of light curves depend on the afterglow parameters. Our afterglow model consists of a 9-dimensional parameter space, which is denoted as $\Theta_{A} = \left \{E_{k},\: n_{0},\: \Gamma_{c},\: \theta_{c},\: \epsilon_{e},\: \epsilon_{B},\: p,\: \theta_{v},\: \: d_{L} \right \}$. 

In order to explore the effect on the SSC light curve in figure \ref{CTA_LC_N}, we first keep the luminosity distance $d_{L}$ and the spectral index $p$ fixed at $250$~Mpc and $2.5$, respectively. We fix the bulk Lorentz factor at the jet axis to be $\Gamma_{c}=650$. We keep $\theta_{c}$ fixed at $8^{\circ}$ and consider two different values of the observer viewing angle $4^\circ$ ($<\theta_c$)  and $10^\circ$ ($>\theta_c$). 
Figure \ref{CTA_LC_N} shows SSC light curves with variations of afterglow parameters. In each of the panels, violet curves represent the $\theta_{v}<\theta_{c}$ case, and dark red curves represent $\theta_{v}>\theta_{c}$ case. In addition, in each panel, we vary one of the afterglow parameters $E_{k}$, $n_{0}$, $\epsilon_{e}$ and $\epsilon_{B}$, between a lower (dashed curve) and a higher value (solid curve). The values taken by the parameters are as follows: $E_{k}$ varies at $10^{50}$ and $10^{51}$ erg; $n_{0}$ having values $10^{-2}$ and $10^{-1}$ $cm^{-3}$,  $\epsilon_{e}$ varies at $10^{-2}$ and $10^{-1}$; and $\epsilon_{B}$ takes values $10^{-3}$ and $10^{-2}$. When one parameter varies, other parameters take fixed values at $E_{k}=10^{50}$ erg, $n_{0}=10^{-1}$ $cm^{-3}$, $\epsilon_{e}=10^{-2}$, $\epsilon_{B}=10^{-3}$. 

Higher $E_k$ and $\epsilon_e$ increase the SSC flux as both increase the seed photon field produced by the non-thermal electrons. A larger $n_0$ corresponds to a larger number of radiating electrons. In the SSC light curve, $\epsilon_B$ does not have a significant effect. The peak time of on-axis observers, occurring at the epoch of deceleration, is directly proportional to $E_k$ and inversely proportional to $n_0$ because the deceleration begins when the inertial mass of the swept-up material equals $E_k/\Gamma_c c^2$. A high $n_0$ corresponds to a larger swept-up mass. 

Here, pair production optical depth and the EBL attenuation factor is calculated for an observer frequency of 250 GeV. The EBL attenuation factor is calculated at luminosity distance $d_{L}=$250 Mpc. TeV photons emitted at this luminosity distance will not be significantly affected by Extragalactic Background Light (EBL) \citep{dominguez2011extragalactic}.

\section{Joint detection with GW detectors} \label{sect4}

The detectability of the inspiral signal is determined by both the sensitivity of the GW detector network and the parameters of the BNS system. The CTA sensitivity, field of view, operational time, and slew timescale, along with the physical parameters of the afterglow, are crucial to determining the detection probability of the associated afterglow in the VHE regime. The most optimistic scenario is when LIGO detects a BNS merger with a localization region well inside the field of view (FoV) of CTA. This leads to a detection or a meaningful upper limit of the VHE signal. On the other hand, if the LIGO localization region is outside or is considerably larger than the CTA FoV, a detection or a meaningful upper limit can not be guaranteed. If the GW localization region is either sufficiently tight or is improved to arc minute scale by the detection of an EM counterpart in timescales of a few seconds, CTA may be able to slew to the location before the VHE counterpart fades below the threshold.

In this section, we explore the joint detection of BNS mergers using LVK and CTA, assuming every BNS merger produces a Gaussian structured jet. We identify the regions of the afterglow parameter space favourable for joint detection and estimate the expected rates of such detections in LIGO cycle O5.

\begin{figure}[ht]
  \centering
   \includegraphics[width=.9\linewidth]{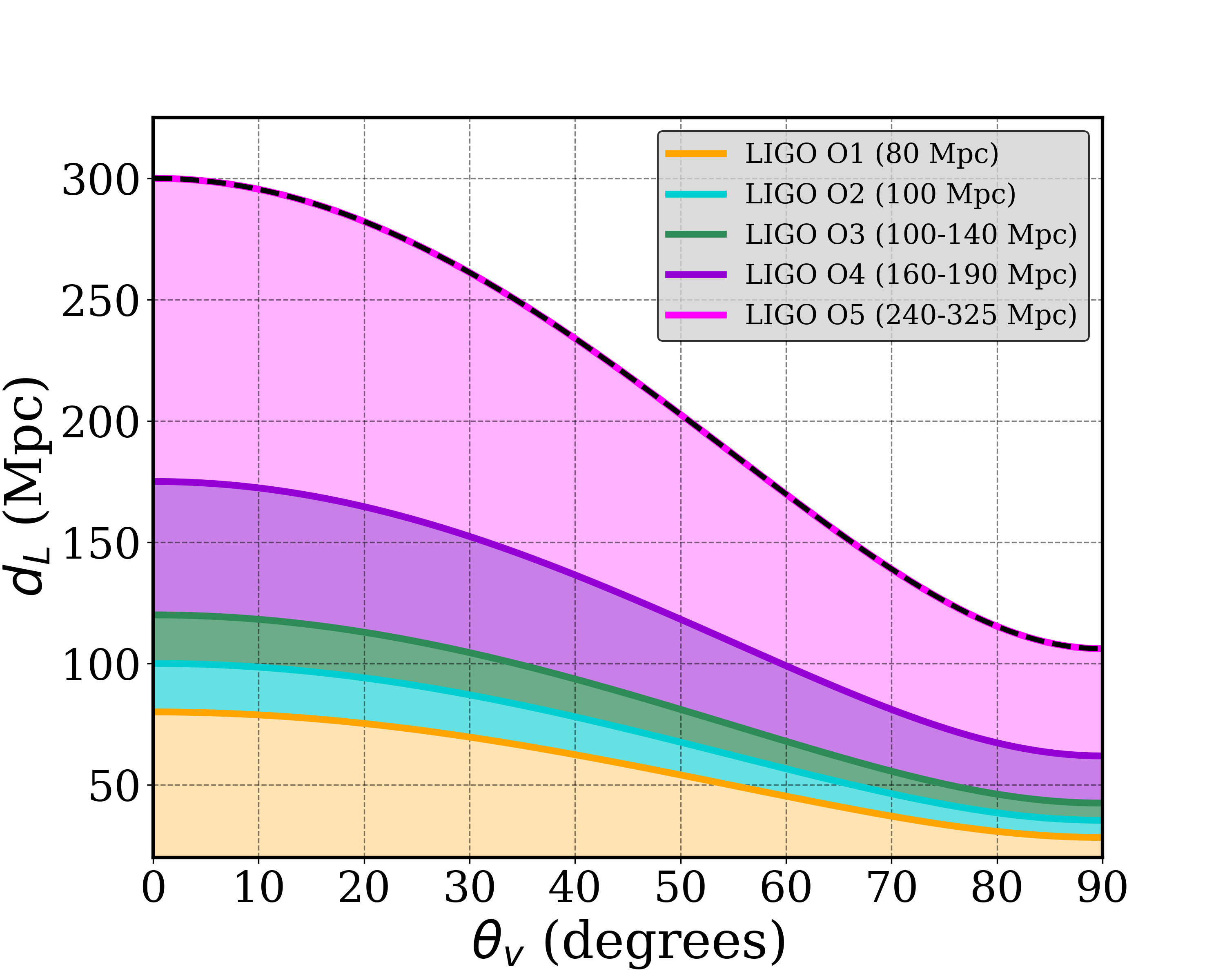}
\caption{Variation of luminosity distance $d_{L}$ with inclination angle $\theta_{v}$ for different observing runs in LIGO: O1 (80 Mpc), O2 (100 Mpc), O3 (100-140 Mpc), ongoing O4 (160-190 Mpc), and upcoming O5 (240-325 Mpc). For each run, the averaged sky position $\bar{H}_{BNS}$ is set to 80 Mpc, 100 Mpc, 120 Mpc, 175 Mpc, and 300 Mpc, respectively, with the orbital inclination angles uniformly distributed in $\cos \theta_{v}$ between [0,1]. The black dotted curve represents the sensitivity of the upcoming LIGO observation run O5, chosen for further analysis of BNS merger event detection. \label{LIGO_obs_run}}
\end{figure}

\subsection{Detection criteria for gravitational wave signals} \label{ligo_only}
For the detection of the chirp signal of the merger with components of $1.4 M_{\odot}$ each, we follow the same prescription used by \cite{duque2019radio}.

The single detector signal-to-noise ratio (SNR) $\rho$ is given by $\rho^2 = \frac{\Omega^{2}_{\rm BNS}}{d_{L}^2}M_{c}^{\frac{5}{3}}S_{I}$ \citep{finn1993observing} where $S_I$ is a function of the detector sensitivity,  $M_c$ is the chirp mass of the binary which equals to $1.22M_{\odot}$ in this case, and $d_L$ is the luminosity distance. $\Omega_{\rm BNS}$ depends on the location and the orientation of the binary orbit in the plane of the sky. 

\citet{duque2019radio} derived an analytical expression for the single detector SNR averaged over all sky positions and orientations of the binary systems. Now, for a given detection threshold, this can be translated to a condition over $\theta_v$ and $d_L$.  Following \citet{duque2019radio}, we consider an SNR of 8 as the detection threshold (i.e., $\rho \ge 8$), which leads to the condition for detection of a merger at a distance $d_L$ and with an orbital inclination angle $\theta_v$, as given below. 
\begin{equation}\label{LIGO_dct_equn}
    d_{L, GW} < H (\theta_{v}) = \left ( \sqrt{\frac{1}{8}} \right )\bar{H}_{BNS} \left ( \sqrt{1+\cos \theta_{v}^{2}+\cos \theta_{v}^{4}} \right ),
\end{equation}
where $H(\theta_{v})$ is the BNS merger horizon and ${\bar{H}_{BNS}}$ is the averaged horizon. 

Using the above equation, in figure \ref{LIGO_obs_run}, we present the expected BNS detection range for LIGO at its different observing runs\footnote{\url{https://www.ligo.caltech.edu/news/ligo20220123}} starting from O1 (80 Mpc), O2 (100 Mpc), O3 (100-140 Mpc), ongoing O4 (160-190 Mpc), and upcoming O5 (240-325 Mpc) run. In the figure, for each of the observing runs, we have taken averaged sky position ${\bar{H}}_{BNS}$ as 80 Mpc, 100 Mpc, 120 Mpc, 175 Mpc, and 300 Mpc, respectively and distributed the orbital inclination angles uniformly in $\cos \theta_{v}$  between [0,1]. Figure \ref{LIGO_obs_run} shows that as $\theta_{v}$ increases, $d_{L,max}$, the farthest distance at which a merger can be detected decreases following the equation \ref{LIGO_dct_equn}. With increased sensitivity in O5, the number of detections of BNS mergers will increase significantly from the ongoing O4 run. To establish BNS merger event detection with LIGO, the upcoming LIGO observation run O5 has been chosen for further analysis, and its sensitivity is represented by the black dotted curve in figure \ref{LIGO_obs_run}.

We simulated a population of $10^{5}$ BNS merger events uniformly distributed in the comoving volume within the range of $30 {\rm Mpc}^3$ to $300 {\rm Mpc}^3$. The viewing angle $\theta_{v}$ is distributed isotropically as $cos\theta_{v} \sim \mathcal{U}(0,1)$. All LIGOs detected events correspond to $d_L(\theta_v) < H(\theta_v)$ satisfying the criterion given in equation \ref{LIGO_dct_equn}.

\begin{figure*}[ht] 
  \centering
    \subfigure[$\theta_{v}$]{\includegraphics[width=.32\textwidth]{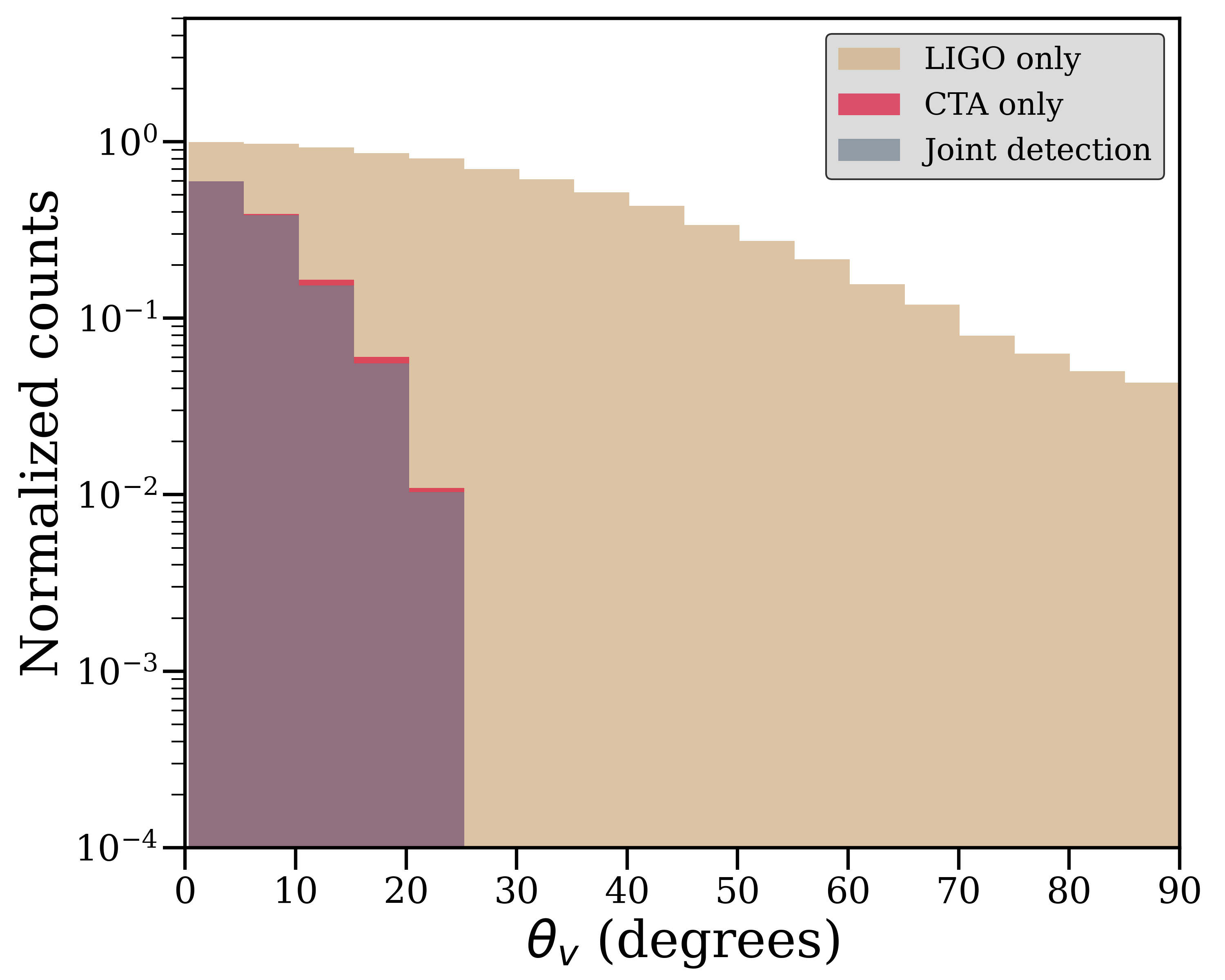}}
  \subfigure[$d_{L}$]{\includegraphics[width=.32\textwidth]{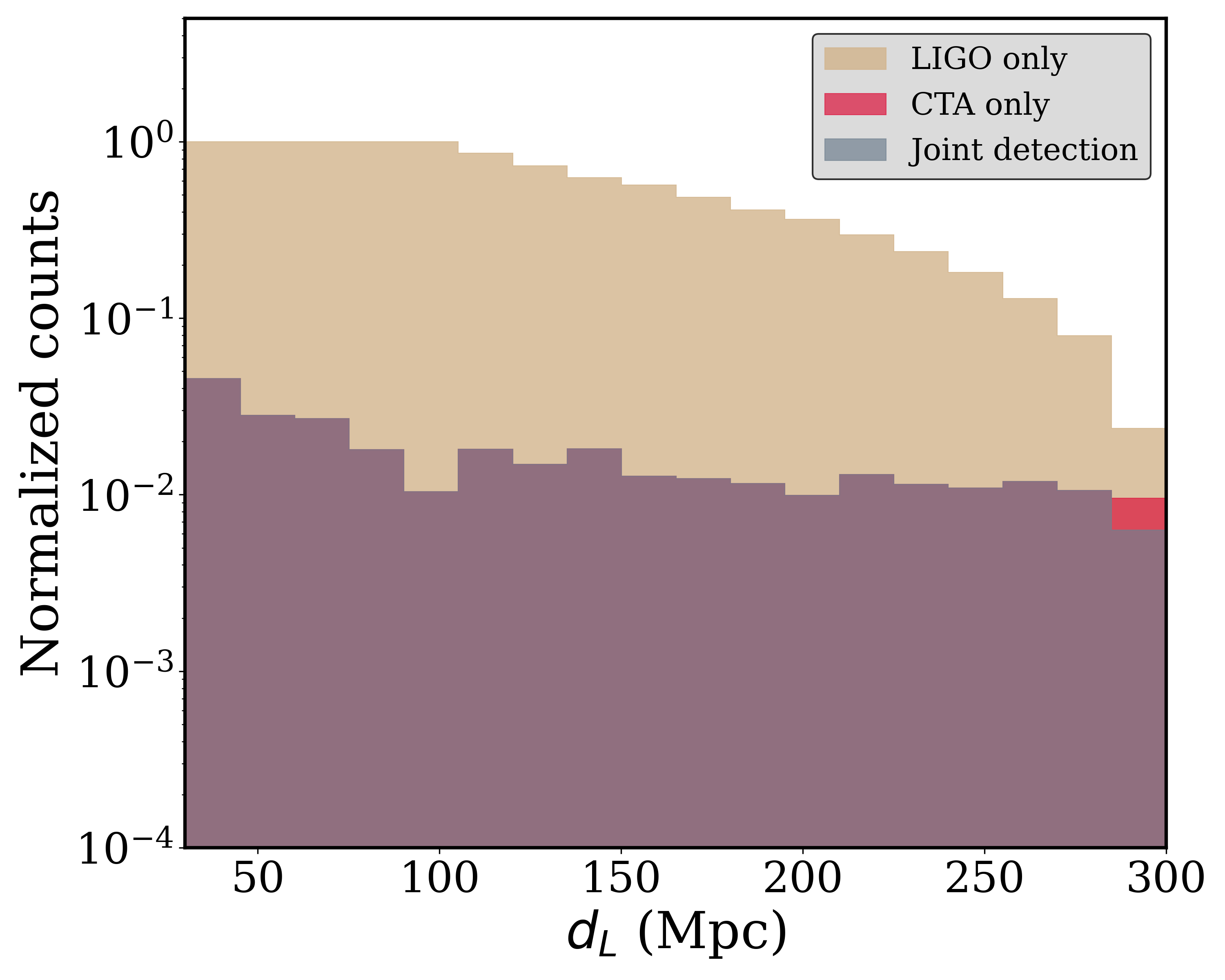}}
  \subfigure[$\theta_{v} / \theta_{C}$]{\includegraphics[width=.32\textwidth]{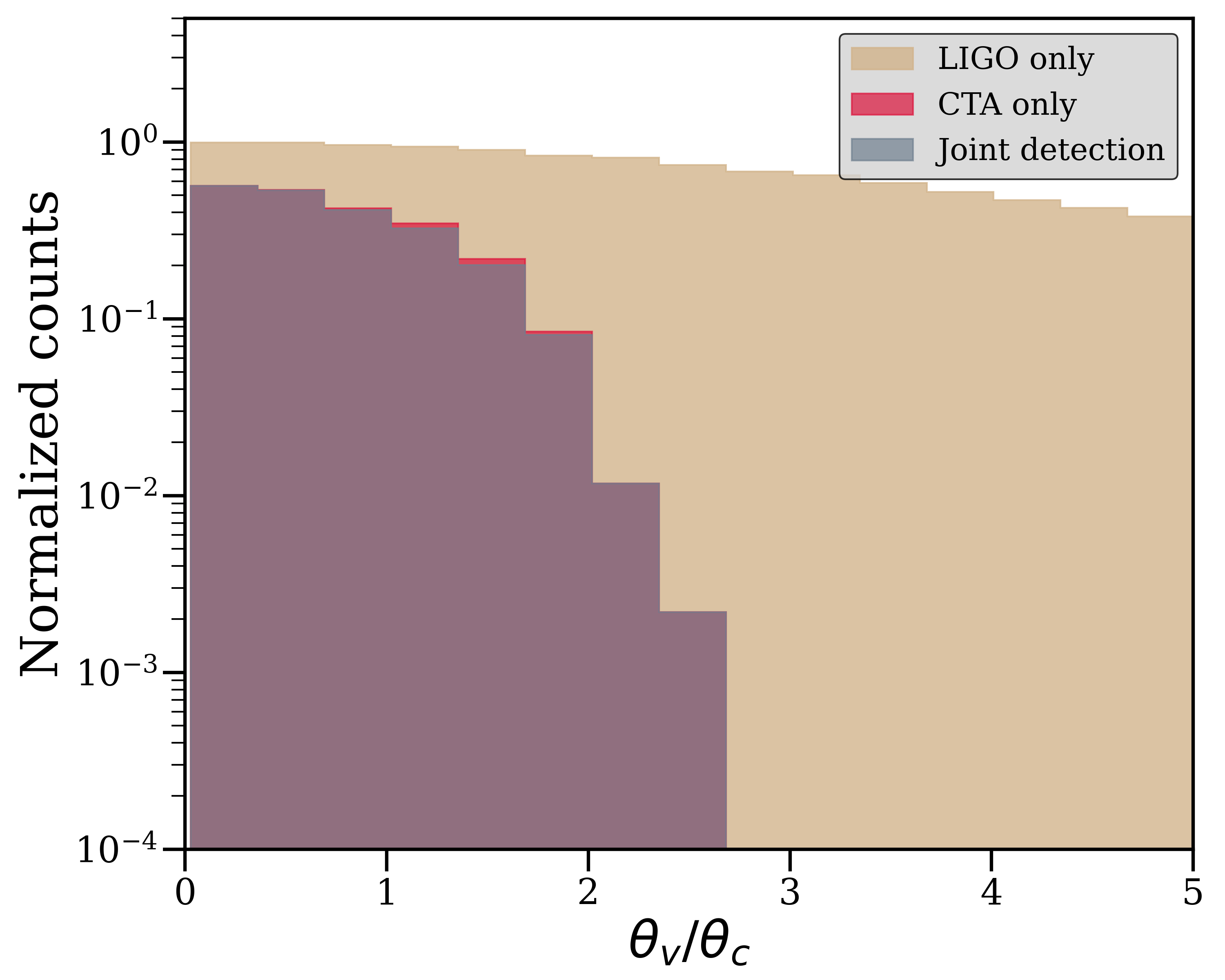}}
\caption{Normalized histograms represent the distribution of events -- LIGO only, CTA only, and joint detected with respect to --- a) $\theta_{v}$, b) $d_{L}$ and c) $\theta_{v} / \theta_{c}$. These plots illustrate the regimes where the values of $\theta_{v}$, $d_{L}$ and $\theta_{v} / \theta_{c}$ are more probable.}
\label{histo_norm_thv_thc_dl}
\end{figure*}

\begin{figure*}[ht] \label{kernel}\label{triangularplot}
    \centering
    \includegraphics[width=\textwidth]{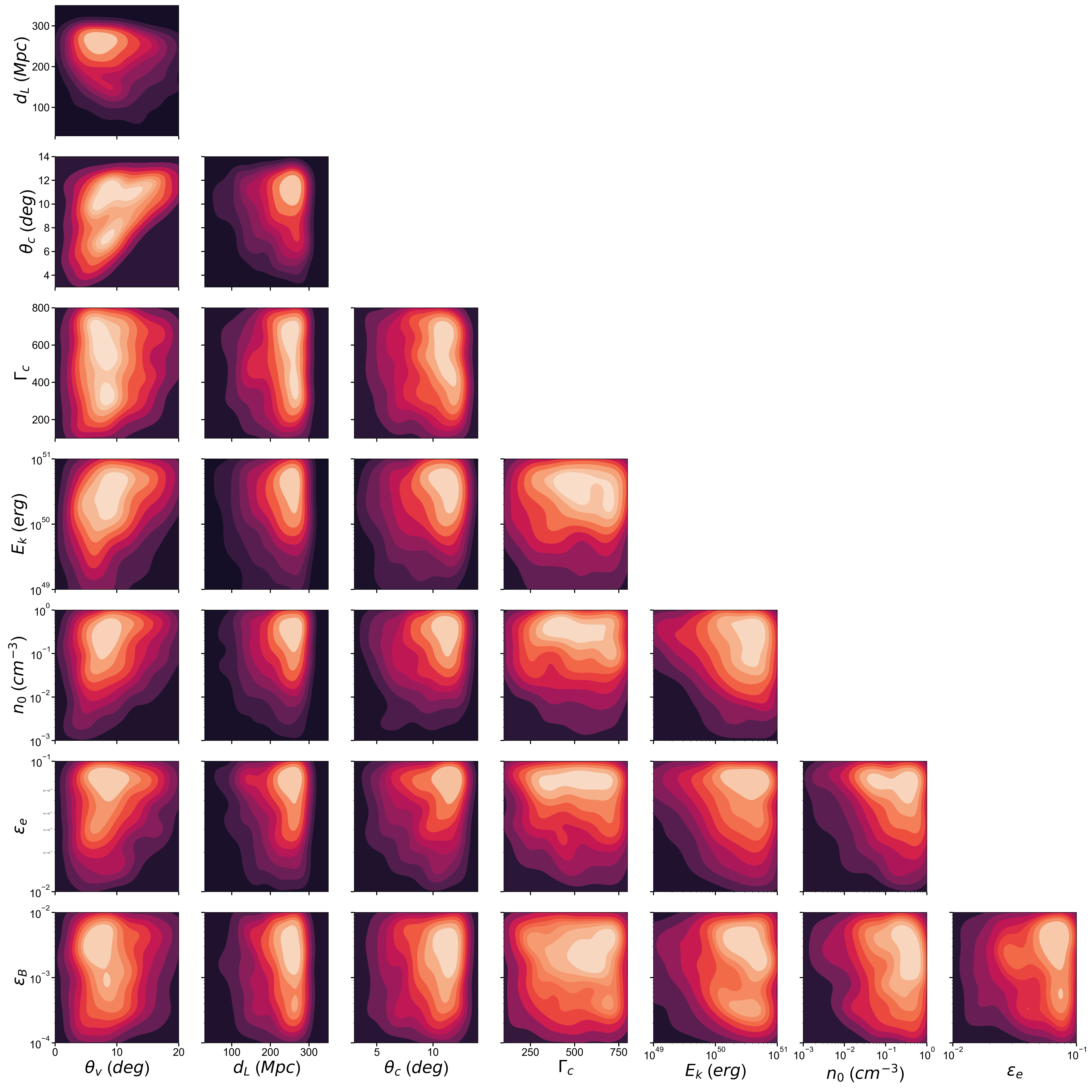}
    \caption{The density plots show correlations within the afterglow parameter space for jointly detected events. These plots illustrate the regimes where all these parameter values are more probable. The brightest regimes indicate the area of higher density, which is the most probable regime jointly detected by CTA and LIGO.}
    \label{param_variation_afterglow_KDE}
\end{figure*}

\begin{figure}[ht]
    \centering
    \includegraphics[width=\linewidth]{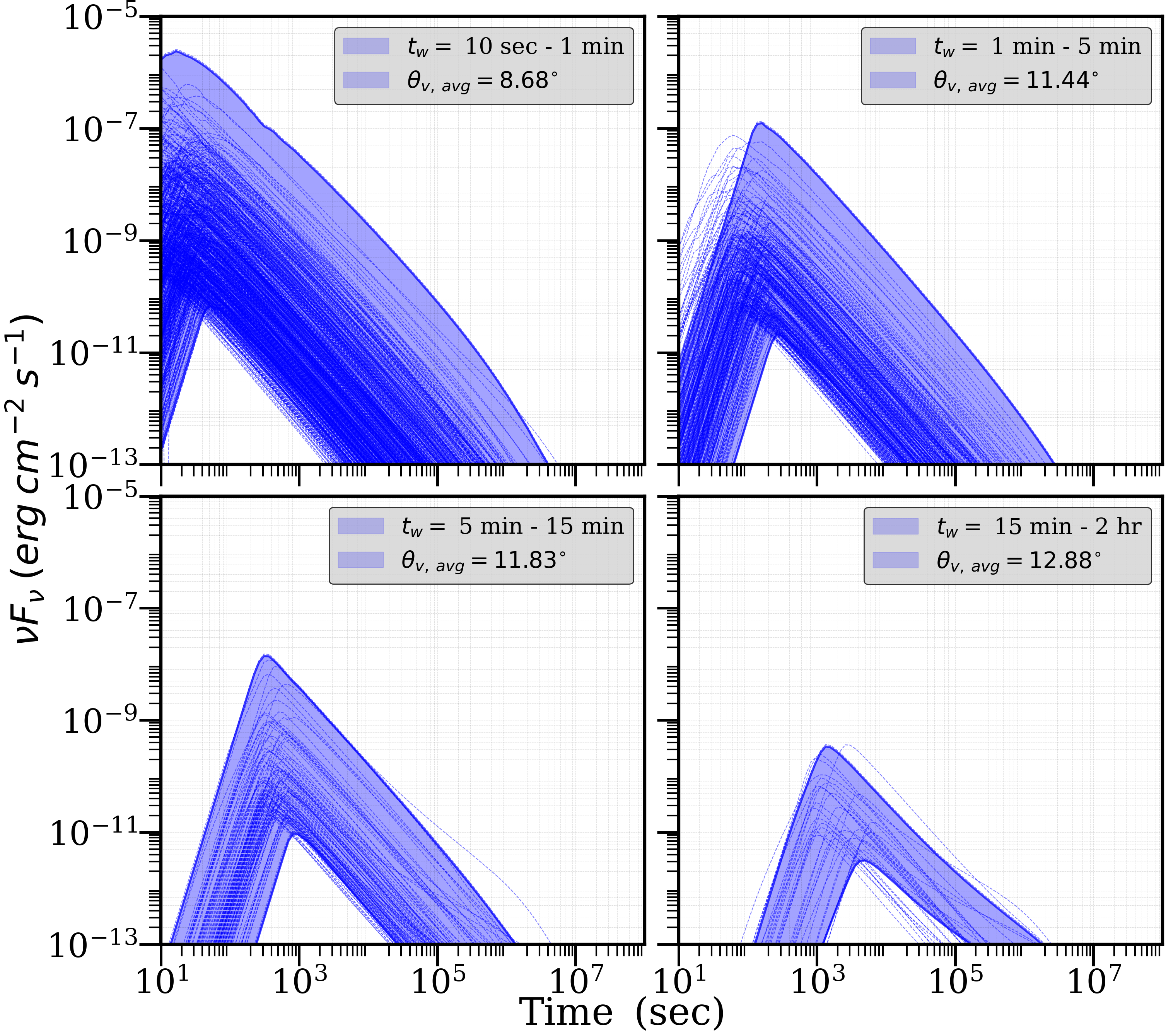}
    \caption{Joint detected light curves are plotted with the variation of peak time from [10 sec, 1 min], [1 min, 5 min], [5 min, 10 min], [10 min, 2hr]. All of the light curves are jointly detected by both LIGO and CTA. The panels show a decrease in the number of joint detected events as the peak time increases, while the average viewing angle, $\theta_{\text{v, avg}}$, increases across the panels with increasing peak time.}
    \label{LC_band}
\end{figure}

\subsection{Joint detection by CTA and LIGO}\label{jd_crit}
Next, we study the joint detection possibilities with LIGO and CTA. We assume that all BNS events are associated with a Gaussian structured jet. The viewing angle $\theta_{v}$ and luminosity distance $d_{L}$ are common for both BNS mergers and GRB jets. Thus, seven afterglow parameters remain, as described in section \ref{CTA-TEV}. For each simulated event, we assign a random value of each afterglow parameter chosen from a prior range consistent with the literature, given in Table \ref{Table1}. We adopt log uniform priors on $E_{k}$, $n_{0}$, $\epsilon_{e}$, and $\epsilon_{B}$, reflecting their wide range of values seen in the literature. $E_{k}$ ranges from $10^{48}$ to $10^{51}$ erg, $n_{0}$ varies between $10^{-3}$ to $1$ $cm^{-3}$, $\epsilon_{e}$ spans from $10^{-2}$ to $10^{-1}$, and $\epsilon_{B}$ ranges between $10^{-4}$ to $10^{-2}$. We varied $\Gamma_{c}$, the bulk Lorentz factor at the jet axis, from $100$ to $800$. The jet core angle $\theta_{c}$ is uniformly distributed between $3^{\circ}$ to $13^{\circ}$. As discussed in section \ref{CTA-TEV}, the spectral index $p$ is fixed to a typical value of $2.5$. 

\begin{table}[ht!] 
\caption{Prior distribution of Gaussian-jet afterglow model\label{Table1}}
\centering
\begin{tabular}{ c c c } \hline \hline

\textbf{Parameter}  & \textbf{Range}           & \textbf{BNS Population} \\ \hline   
$E_{k}$ (erg)       & $10^{49}-10^{51}$        & uniform log prior       \\ 
$n_{0}$ (cm$^{-3}$) & $10^{-3} - 1$            & uniform log prior       \\ 
$\epsilon_{e}$      & $10^{-2}-10^{-1}$        & uniform log prior       \\ 
$\epsilon_{B}$      & $10^{-4}-10^{-2}$        & uniform log prior       \\ 
$\theta_{v}$ (deg)  & $0^{\circ}$-$90^{\circ}$ & uniform in $\cos[0,1]$  \\ 
$\theta_{c}$ (deg)  & $3^{\circ}$-$13^{\circ}$ & uniform prior           \\ 
$\theta_{j}$ (deg)  & $25^{\circ}$             & fixed                   \\ 
$\Gamma_{c}$        & 100-800                  & uniform prior           \\ 
$d_{L}$ (Mpc)       & 30-300                   & uniform prior           \\ 
$p$                 & 2.5                      & fixed                   \\ \hline 
\end{tabular}
\end{table}

As mentioned in section \ref{LCs}, we have used the differential flux sensitivity of the CTAO-Northern array at an energy of 250 GeV to determine the detection threshold for the simulated light curves, with an integration time ranging from a few seconds to a few hours. We consider an event to be a detection if its light curve crosses the detection threshold of CTA at any point in time. Following equation \ref{LIGO_dct_equn}, we evaluate whether the $d_L$ and $\theta_v$ of the event meet LIGO's detection criteria. The events detected by CTA are labelled as CTA only, and detected by LIGO are labelled as LIGO only. The events satisfying both the CTA and LIGO detection criteria are classified as jointly detected. In the next section, we will discuss the properties of the parameter space favouring joint detection.

\subsection{Characteristics of Joint detection events}\label{joint_dect_sec}

We obtained the detected events using the criteria outlined in Sections \ref{LCs}, \ref{ligo_only}, and \ref{jd_crit}. To understand how joint detection probability is affected by viewing angle and distance, we plotted normalized 1D histograms of $\theta_v, d_L,$ and $\theta_v/\theta_c$ (figure \ref{histo_norm_thv_thc_dl}). While $\theta_v$ plays a crucial role in LIGO detectability, the beaming effect governing afterglow detection depends on the observer’s misalignment relative to the jet core angle. Therefore, the ratio $\theta_v/\theta_c$ is important in addition to the individual value of $\theta_v$. Since the number of simulated events is distributed uniformly across $\cos{\theta_v}$  and $d_L^3$, rather than $\theta_v$ and $d_L$ themselves, in the 1D normalized histograms (figure \ref{histo_norm_thv_thc_dl}), we implemented a bin-wise normalization to represent the distributions accurately. For every distribution, we normalized the number of detected events in each bin by the total number of simulated events in that bin to accurately reflect the true dependence.

\begin{widetext}
\begin{table*}[ht]
\caption{Details of all simulated BNS merger events and detection numbers for different observer viewing angle ($\theta_v$) ranges.\label{detection_rate_table1}}

\begin{tabular}{c c c c}
\hline \hline
\textbf{Observer viewing angle} & \textbf{All BNS Merger events} & \textbf{LIGO detected events} & \textbf{Joint detected events} \\ \hline
$0^{\circ}$ to $10^{\circ}$  & 1463  & 1432 ($97.88 \%$)  & 654 ($44.7 \%$)  \\ 
$10^{\circ}$ to $15^{\circ}$ & 1863  & 1732 ($92.96 \%$)  & 303 ($16.26 \%$) \\ 
$15^{\circ}$ to $20^{\circ}$ & 2715  & 2348 ($86.48 \%$)  & 159 ($5.86 \%$)  \\ 
$20^{\circ}$ to $30^{\circ}$ & 7564  & 5705 ($75.42 \%$)  & 39 ($0.52 \%$)   \\ 
$30^{\circ}$ to beyond       & 86395 & 17997 ($20.83 \%$) & 0 ($0 \%$)       \\ \hline
\end{tabular}
\end{table*}
\end{widetext}

The 1D histogram shows that the probability of joint detection is highest for the bin $0^{\circ} \le \theta_{v} \le 5^{\circ}$. Although the number of simulated events increases as $\theta_v$ and $d_L$ increase,  events at greater distances can be detected by LIGO at smaller viewing angles because the maximum detectable distance decreases as $\theta_v$ increases (see Figure \ref{histo_norm_thv_thc_dl}). In the electromagnetic case, smaller viewing angles (more precisely, smaller $\theta_v/\theta_c$) allow the observer to see higher flux levels due to the direct visibility of the central bright jet core, where relativistic beaming is more pronounced, leading to a higher fraction of detections. Consequently, both for LIGO and CTA, the fraction of detected events is higher at smaller viewing angles. Due to the beaming effect, CTA detection sharply drops as $\theta_v$ increases, and no events with viewing angles ($>25^{\circ}$) are detected by CTA. Therefore the joint detection is driven by CTA, which is also visible from both the histograms and the detection percentage given in Table \ref{detection_rate_table1}. As a result, only a $\sim 4$ percentage of LIGO-detected events are detected by CTA. The observer is within the jet core either due to a large core angle or a small viewing angle. The histogram of $\theta_{v}/\theta_{c}$, which is applicable only to electromagnetic detections, shows that a $\theta_v/\theta_c \lesssim 2.7$ is required for an event to be detected by CTA.

To examine the correlations among the physical parameters that favour joint detection, we generated 2D kernel density plots of every pair of the 8 parameters $\theta_v$, $d_L$, $\theta_c$, $\Gamma_c$, $E_k$, $n_0$, $\epsilon_e$, and $\epsilon_B$ (figure \ref{param_variation_afterglow_KDE}). In these density plots, the colour transition from darker to lighter shades indicates an increase in the population of jointly detected events.

The density plots in figure \ref{param_variation_afterglow_KDE} reveal correlations within the afterglow parameter space of joint-detected events. Higher values of $E_k$, $n_0$, and $\epsilon_e$ result in increased VHE flux, as described in section \ref{LCs} and illustrated in figure \ref{CTA_LC_N}. This enhances the likelihood of CTA detection and, consequently, joint detection also. Most joint detected events have values above $10^{50}$ erg for $E_k$, $10^{-1}$ $cm^{-3}$ for $n_0$ and $6 \times 10^{-2}$ for $\epsilon_{e}$. The threshold values are influenced by the distribution of $\theta_c$; a smaller $\theta_c$ leads to a greater reduction in the flux of misaligned events, necessitating even higher values of $E_k, n_0$, and $\epsilon_e$ for the VHE afterglow to be detected.

Additionally, we observe that higher values of $\Gamma_{c}$, ranging from $550$ to $750$, are more likely in joint detection scenarios, as a higher bulk Lorentz factor significantly enhances the flux at the light curve peak, which results from the deceleration of the fireball. However, $\Gamma_c$ shows no strong correlation with other parameters as the effect is not as pronounced compared to $E_k$, $n_0$, and $\epsilon_e$. The same applies $\epsilon_B$ as shown in Figure 1. Although higher $\epsilon_B$ increases the magnetic field and, consequently, the seed photon (synchrotron) flux, its impact on the SSC flux is not as significant as that of $E_k, n_0$, and $\epsilon_e$. 

As mentioned earlier, a lower $\theta_v/\theta_c$ favours EM detection leading to a strong anticorrelation between $\theta_c$ and $\theta_v$ as seen in figure \ref{param_variation_afterglow_KDE}. A preference for larger jet core angles ($\theta_{c} \sim 10^{\circ}-12^{\circ}$) is observed among the jointly detected events.

While density plots offer insights into parameter dependencies, understanding the variation of peak flux and peak time necessitates plotting light curves of the jointly detected events. This approach also helps to understand the CTA slew timescale required for detection. 

\begin{table}[ht]
\caption{Number of detected light curves based on peak time.\label{detection_rate_tw}}
\resizebox{\columnwidth}{!}{%
\begin{tabular}{c c}
\hline \hline
\textbf{Observer time window} & \textbf{Joint detected events} \\ \hline  
10 sec to 1 min               & 699                            \\ 
1 min to 5 min                & 318                            \\ 
5 min to 15 min               & 107                            \\ 
15 min to 2 hr                & 31                             \\ \hline
\end{tabular}%
}
\end{table}

In figure \ref{LC_band}, we categorize all light curves into four panels based on their peak times as follows: I) $10$ seconds to $1$ minute, II) $1$ minute to $5$ minutes, III) $5$ minutes to $15$ minutes, and IV) $15$ minutes to $2$ hours. The faint blue dotted lines are the light curves detected jointly by CTA and LIGO and the favourable condition for detection will occur if CTA gets a trigger during the rising phase of the light curve. Thus, it will acquire the GRB within its typical slew time of 20 sec \citep{banerjee2023pre} and will have enough time to detect the source. Table \ref{detection_rate_tw} presents the number of joint detected events across different observational time windows and reveals that the number of joint detected events decreases as the peak time increases. This is because the peak time and peak flux are influenced by $\theta_v$; a larger $\theta_v$ results in a delayed peak and reduced peak flux. As the number of jointly detected events decreases with increasing $\theta_v$ (see table \ref{detection_rate_table1}), the upper leftmost panel contains the largest number of light curves, while the bottom rightmost panel contains the fewest. Consequently in figure \ref{LC_band}, the average observer viewing angle, $\theta_{\text{v, avg}}$ increases from panels I to IV with values $8.68^{\circ}$, $11.44^{\circ}$, $11.83^{\circ}$, and $12.88^{\circ}$ respectively.

\begin{figure*}[ht]
    \centering
     \includegraphics[width=0.85\textwidth]{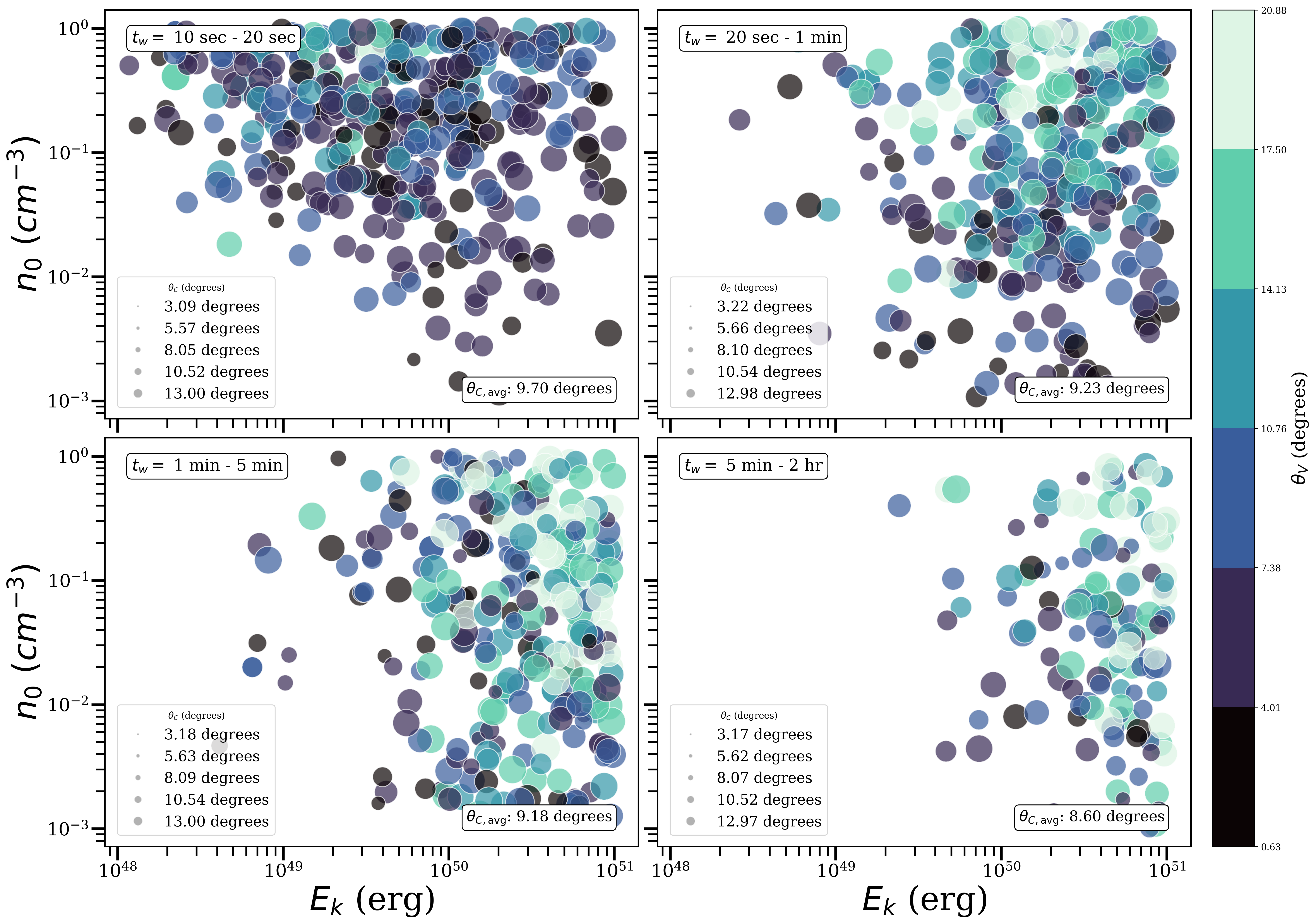}
    \caption{The dependencies of peak time on the parameters $E_k$, $n_0$, $\theta_v$, and $\theta_c$ were analysed across four-time windows: (I) 10-20 seconds, (II) 20 seconds-1 minutes, (III) 1-5 minutes, and (IV) 5 minutes-2 hour. We observed that at lower peak times, there are a notable number of events with higher values of $n_0$ and the probability of getting higher $n_0$ values decreases as peak time increases. Similarly, at lower peak times, there are significant numbers of events with lower $E_k$ values, and the density of these lower $E_k$ events also decreases with increasing peak time. Additionally, as peak time increases, the colour axis representing $\theta_v$ shifts to lighter shades or higher values and the sizes of the circles representing $\theta_{c}$ decreases, thus also the value of $\theta_{c, avg}$ decreases across the panels.}\label{CTA_tpeak}
\end{figure*}

In Fig. \ref{CTA_tpeak}, we examined the dependency of peak time on the parameters $E_k$, $n_0$, $\theta_v$, and $\theta_c$ across four-time windows: I) $10-20$ seconds, II) $20$ seconds to $1$ minute, III) $1$ minute to $5$ minutes and IV) $5$ minutes to $2$ hours for joint detected BNS merger events. The figure agrees with the peak time dependency on various afterglow parameters, which is discussed in section \ref{LCs} in detail. Since more mass is required to decelerate a higher energy fireball, the peak time, which is decided by the epoch of deceleration if $\theta_c <\theta_v$, increases as $E_k$ increases and $n_0$ decreases. The peak is delayed till the observer enters the beaming cone of the core of the jet if $\theta_c <\theta_v$, and is greater for higher values of $\Gamma (t_{\rm obs})$ and $\theta_v/\theta_c$. Since $\Gamma (t_{\rm obs}) \propto (E_k/n_0)^{1/3}$, the peak time increases for higher $E_k$ and lower $n_0$. As discussed earlier, due to higher observer misalignment, higher $\theta_v$ or smaller $\theta_c$ lead to a delayed peak. For the same reason, that $\theta_{c, avg}$ decreases as peak time increases. In our population analysis through density plots in figure \ref{param_variation_afterglow_KDE}, we observe that events having lower values of $\theta_{v}$, higher values of $\theta_{c}$, $E_{k}$, and $n_{0}$ are most probable for joint detection. This result leads us to anticipate that the majority of light curves will fall into the early peak time panel which is also observed in figure \ref{LC_band}.

\subsection{Joint detection rate of BNS mergers} \label{event_rate}
In this section, we estimate the joint sGRB-GW detection rate for CTA with the upcoming LIGO O5 run. We simulated $10^5$ pairs of $d_L$ and $\theta_v$ sampled from  distributions defined by $d_L^3 \sim \cal{U}$$(30^3,300^3)$ and $\cos(\theta_v) \sim U(0,1)$. Each pair represents a BNS merger. The LIGO detection criterion given in equation \ref{LIGO_dct_equn} assumes component masses of $1.4 M_{\odot}$ each and uses a single detector SNR threshold of 8. 

Assuming all BNS mergers are associated with Gaussian structured jets similar to GRB170817A and that CTA can slew to cover the merger location in its FoV before the flux drops below the sensitivity limit, we calculated the joint detection probability. Of the simulated events, 29,214 (29\%) are detectable by LIGO in O5, while 1,155 (1\%) are detectable jointly by both CTA and LIGO. These fractions are highly sensitive to assumptions about the distributions of afterglow physical parameters.

Given a BNS merger rate of $10-1700$ Gpc$^{-3}$ yr$^{-1}$ \citep{abbott2023population}, these numbers translate to a LIGO detection rate of $(0.08-13)$ BNS mergers per year within $300$~Mpc during O5 and a joint detection rate of $(0.003-0.5)$ events per year with CTA. 

\section{Discussion and Conclusion} \label{Conclusion}

In this study, we describe the detection prospects of sGRB-GW events with CTA in synergy with the upcoming LIGO O5 observation run. To estimate the joint detection rate of BNS merger events, we have considered the Gaussian structured jet profile in our calculation for predicting off-axis emission. Our model explains VHE afterglow off-axis emission (above 100 GeV) from sGRBs through the SSC emission process while accounting for Klein-Nishina correction, internal absorption due to pair production and EBL correction factors to further modify the SSC flux at the VHE regime. We have simulated $10^5$ BNS merger events to explore the joint-detection prospects of sGRB-GW events. Using the differential flux sensitivity of CTAO at $250$ GeV and the improved sensitivity of upcoming LIGO O5, we systematically examine the afterglow parameter space favourable for joint detection. Moreover, we examine the influence of different afterglow parameters on TeV detections through SSC light curves in CTA observations. We further calculate the joint detection rates of BNS merger events for CTA in synergy with the LIGO O5 run.

The synergistic study of LIGO with other $\gamma-$ray detectors has been well explored in previous few studies. \citet{howell2019joint} presented a detailed analysis of the prospect of events jointly detected by Fermi-GBM with LIGO O3 run and found a joint event rate of 4 $yr^{-1}$ in this detection scenario. \citet{bhattacharjee2024joint} also explores the prospect of joint detection from BNS merger using current and upcoming GW facilities IGWN4 and IGWN5 with existing GRB satellites (Fermi and Swift). With the proposed Daksha mission, the rate of joint detection will significantly increase 2-9 times than existing. \citet{bartos2019gravitational} investigates CTA follow-up observations of GW-detected GRBs in the TeV energy range, showing that CTA could detect VHE flux from events like GRB090510, even with a one-hour delay after a BNS merger detection. They also showcased that CTA could observe a fainter event, up to $10^{-2}$ dimmer than GRB090510, with a 20-minute delay for sources around 500 Mpc. While this manuscript was being prepared, \citet{pellouin2024very} also proposed a comprehensive numerical model for simulating the SSC afterglow of structured jets in BNS across the radio to very high energy, which they applied to explain high energy emission from GW 170817. They reported a flux below HESS's upper limits and proposed that the upcoming CTA could detect such events if the jet is slightly off-axis, in a high-density medium, and within 100-400 Mpc.

Our approach to model sGRB afterglow SSC emission showcases few similarities with the proposed model of \citet{pellouin2024very}. Although, as mentioned earlier, through our model, we aim to study the joint detection prospect of BNS merger events with CTA and LIGO upcoming O5 run. Our analysis reveals that extreme off-axis viewing angles $(> 30^{\circ})$ are generally not detectable in the joint detection, while parameters such as high jet core angle $\theta_c$ $(10^{\circ}-12^{\circ})$, kinetic energy $E_k$ $(> 10^{50}$ erg) and ambient number density $n_0$ $(> 10^{-1}$ cm$^{-3})$ enhance the prospect of joint detection. The ratio of $\theta_v / \theta_c$ emerges as a critical parameter, where we find that merger events are not likely to be detected if the ratio is $>2.7$. Additionally, high values of $\epsilon_e$, $\epsilon_B$ and $\Gamma_c$ are more likely to get jointly detected. Our findings indicate that detection scenarios are highly sensitive to the ranges and distributions of these physical parameters. Our observations further showcase that light curves peaking at early times (within 1 minute) are more likely to be detected among all jointly detected events. As we shift to a later peak time window, both the number of detectable light curves and their peak fluxes decrease. Because, at this later peak time window, the average $\theta_v$ of light curves increases gradually, making the light curves peak at a delayed time due to less pronounced relativistic beaming effect for $\theta_{v}>\theta_{c}$. The peak time also showcases dependencies on $\theta_c$, $E_k$, $n_0$, where high $n_0$ and $\theta_c$ push the peak time earlier and high $E_k$ shifts the peak time later. For a local BNS merger rate of $10-1700$ ${\rm Gpc}^{-3} {\rm yr}^{-1}$, we found the predicted LIGO detection rate of $(0.08-13)$ BNS mergers per year and a joint detection rate of $(0.003-0.5)$ events per year with CTA. Our analysis framework thus reveals that the joint detections with CTA and LIGO will enhance understanding of the BNS merger events with greater details and play a crucial role in paving the path for advancements in multimessenger astronomy and the emerging era of gravitational wave and gamma-ray observations.

\section*{Acknowledgments}
{T. Mondal and S. Chakraborty acknowledge the support of the Prime Minister's Research Fellowship (\href{https://www.pmrf.in/}{PMRF}). T. Mondal extends special thanks to Prof. Sonjoy Majumder from the Department of Physics at IIT Kharagpur for his invaluable guidance. The authors would also like to acknowledge \href{http://www.hpc.iitkgp.ac.in/}{Paramshakti Supercomputer facility} at IIT Kharagpur—a national supercomputing mission of the Government of India, for providing the necessary high-performance computational resources.}

\bibliography{GRB_LIGO}{}
\bibliographystyle{aasjournal}

\end{document}